\documentclass[twocolumn,prd,amsmath,amssymb,floatfix,nofootinbib]{revtex4}

\usepackage[colorlinks=true,urlcolor=black]{hyperref}

\hypersetup{backref = true,pagebackref = true,hyperindex = true, colorlinks = true,
  breaklinks = true, urlcolor = blue, linkcolor = blue, bookmarks = true,
  bookmarksopen = true, citecolor=red}

\usepackage{bm}
\usepackage{color}
\usepackage{dcolumn}
\usepackage{graphics}
\usepackage{graphicx}
\usepackage{subfigure}
\usepackage{slashed}
\usepackage{placeins}

\allowdisplaybreaks
\usepackage{pdfpages}
\usepackage{eso-pic}
\usepackage{everyshi}
\usepackage{pdflscape}

\def\al{\alpha}

\def\veps{\varepsilon}

\newcommand{\ep}{\varepsilon}
\newcommand{\oep}{\overline{\varepsilon}}

\def\be{\begin{equation}}
\def\ee{\end{equation}}
\def\bea{\begin{eqnarray}}
\def\eea{\end{eqnarray}}
\def\bse{\begin{subequations}}
\def\ese{\end{subequations}}
\def\bc{\begin{center}}
\def\ec{\end{center}}

\def\ra{\rightarrow}

\def\nonum{\nonumber}

\def\I{{\rm i}}

\def\D{{\rm d}}
\def\Ord{{\rm O}}

\newcommand{\ie}{{\it i.e.}}
\newcommand{\eg}{{\it e.g.}}

\begin{document}

\title{Landau-Khalatnikov-Fradkin transformation \\ in three-dimensional quenched QED}
       \author{
       V.\ P.~Gusynin$^{1}$, A.\ V.~Kotikov$^{2}$ and S.~Teber$^{3}$
       }
\affiliation{
$^1$Bogolyubov  Institute  for  Theoretical  Physics,  Kyiv 03143,  Ukraine.\\
$^2$Bogoliubov Laboratory of Theoretical Physics, Joint Institute for Nuclear Research, 141980 Dubna, Russia.\\
$^3$Sorbonne Universit\'e, CNRS, Laboratoire de Physique Th\'eorique et Hautes Energies, LPTHE, F-75005 Paris, France.
 }

\date{\today}

\begin{abstract}
We study the gauge-covariance of the massless fermion propagator in  three-dimensional
quenched Quantum Electrodynamics in the framework of dimensional regularization in  $d=3-2\ep$.
Assuming the finiteness of the quenched perturbative expansion, that is the existence of the limit $\ep \to 0$, we state that, exactly in $d=3$, all
odd perturbative coefficients, starting with the third order one, should be zero in any gauge.
\end{abstract}

\maketitle

\section{Introduction}
\label{Sec:Introduction}

Quantum electrodynamics in three space-time dimensions (QED$_3$) with $N$ flavors of four-component massless
Dirac fermions has been attracting continuous attention as a useful field-theoretical model for the last forty years. It served as a toy model
to study several key quantum field theory problems such as infrared singularities in low-dimensional theories with massless
particles, non-analyticity in coupling constant, dynamical symmetry breaking and fermion mass generation, phase transition
and relation between chiral symmetry breaking and confinement.

In the last {three} decades, QED$_3$ found many applications in condensed matter physics, in particular, in high-$T_c$ superconductivity
\cite{Dorey1992,Franz2001,Herbut2002}, planar antiferromagnets \cite{Farakos1998}, and graphene \cite{Semenoff1984} where quasiparticle
excitations have a linear dispersion at low energies and are described by the massless Dirac equation in $2+1$ dimensions (for graphene,
see reviews in Ref.~\cite{reviews}).

QED$_3$ is described by the action (in Euclidean formulation)
\bea
S=\int d^3x\left[\frac{1}{4}F^2_{\mu\nu}+\bar\Psi_i\gamma_\mu D_\mu\Psi_i\right],
\eea
where $D_\mu=\partial_\mu-ieA_\mu$, $i=1,2,\dots N$, Euclidean gamma matrices satisfy $\gamma^\dagger_\mu=\gamma_\mu$,
$\{\gamma_\mu,\gamma_\nu\}=\delta_{\mu\nu}$, and the gauge coupling constant has a dimension $[e]=(\mbox{mass})^{1/2}$.
It is super-renormalizable with a mass scale $e^2 N/8$. As such, the model is plagued with severe infrared (IR) singularities in the loop
expansion in $e^2$ of various Green's functions since, for dimensional reasons, higher-order diagrams contain higher powers of momentum in
the denominator. For example, the fermion propagator is affected by IR divergences starting at two-loops. The problem of IR divergences in QED$_3$
has been intensively investigated by various methods in the 80s \cite{Jackiw1981,Templeton1981,Bergere:1982br,Guendelman1983,King:1985hr}.

In order to better appreciate how these IR singularities may be cured, let's recall that the effective dimensionless coupling of QED$_3$ may be written as:
\bea
\overline{\al}(q) = \frac{e^2}{q\,(1 + \Pi(q^2))} = \left\{
    \begin{array}{ll}
            e^2/q \quad & \quad q \gg e^2 N/8 \\
            8/N  \quad & \quad q \ll e^2 N/8 \, ,
            \end{array}
        \right.
\eea
where the one-loop polarization operator was used: $\Pi(q^2) =  N e^2 / (8q)$. The corresponding beta function reads:
\be
\beta(\overline{\al}) = q\,\frac{\D \overline{\al}(q)}{\D q} = - \overline{\al}\,\left( 1 - \frac{N}{8}\,\overline{\al} \right) 
\label{beta-albar}
\ee
and displays two stable fixed points: an asymptotically free UV fixed point ($\overline{\al} \ra 0$) and an interacting IR fixed point ($\overline{\al}
\ra 8/N$).\,\footnote{{At the next-to-leading order of the $1/N$-expansion, the polarization operator reads: $\Pi(q^2) =  N C e^2 / (8q)$
where $C = 1 + (184/9\pi^2 - 2)/N$. This result was obtained independently and by different methods in
Refs.~[\onlinecite{Gracey:1993iu,Gusynin2003,Teber:2012de}].
At the interacting IR fixed point, the beta function therefore retains the form (\ref{beta-albar}) with $N \ra N C$.
This in turn implies that the IR fixed point remains stable as it is only weakly affected by next-to-leading order corrections:
$\overline{\al} \ra 8/(NC) \approx (8/N)(1-0.071/N)$, in agreement with Eq.~(30) of [\onlinecite{Gusynin2003}]. To the best of our knowledge,
calculations at the order $1/N^3$ have not been performed yet.}}

The presence of an infrared fixed point in QED$_3$ is intriguing especially because of the possible existence of an analogous
fixed point in four-dimensional $SU(N_c)$ gauge theories with $N$ massless fermions. In this case, there exists a so-called conformal window: a region of
values of colors $N_c$ and flavors $N$, for which the beta function has a form resembling Eq.~(\ref{beta-albar}). Thus, the theory is asymptotically free
at short distances while the long distance physics is governed by a non-trivial fixed-point \cite{Banks1982} (for studies on the lattice and an extensive list
of references see Ref.~\cite{Debbio2011}).

At the fixed points of the beta-function (\ref{beta-albar}), IR singularities of QED$_3$ are cured.
In the IR limit, $q\ll e^2N/8$, this goes through an infrared softening of the photon propagator,
due to vacuum-polarization insertions, together with the disappearance of the dimensionful $e^2$
in favour of a dimensionless coupling constant proportional to $1/N$.
The theory has then the same power counting as the (renormalizable) four-dimensional one.
In the following, we shall refer to this limit as the large-$N$ limit of QED$_3$.
In this limit, the fermion propagator is affected by (standard) UV singularities that can be renormalized \cite{Appelquist1986}.

Massless QED$_3$ also plays an important role in studying the problems of dynamical symmetry breaking and fermion mass generation in gauge
theories. This comes from the fact that the properties (\ref{beta-albar}) are reminiscent of those of QCD: the intrinsic dimensionful parameter
$e^2N/8$ plays a role similar to the QCD scale $\Lambda_{\rm QCD}$, and the effective coupling $\overline{\al}(q)$ approaches zero at large momenta $q$.
The main question that has been debated for a long time is then whether  there exists a critical fermion flavor number, $N_{cr}$,
separating the chiral-symmetric and the chiral-symmetry-broken phases \cite{Pisarski1984,Appelquist1986,Appelquist1988,Nash1988,Atkinson1990,Pennington1991,Kotikov1993,GHR-1996,Maris1996, Gusynin2003b,Fischer2004}
(for recent studies in this direction see, \eg, Refs.~\cite{Gusynin:2016som,Kotikov:2016wrb,Kotikov:2016prf,Kotikov:2020slw,Karthik2019}).

Analytical studies of chiral-symmetry breaking and mass generation in QED$_3$ are based on using truncated Schwinger-Dyson (SD) equations with some
approximations (ansatzes) for the full fermion-photon vertex. For example, the simplest one is the replacement
of the full vertex by the bare vertex $\gamma_\mu$ (the ladder approximation). A more sophisticated way is to use
the vertex consistent with the vector Ward-Takahashi identity and satisfying several other requirements such as the absence of kinematic poles
and multiplicative renormalizability. The most known among them are the Ball-Chiu \cite{Ball1980} and the Curtis-Pennington \cite{Curtis1990} vertices
constructed for the quenched approximation of four-dimensional QED (QED$_4$) -- for similar vertices in QED$_3$, see \cite{Bashir:2000rv}.

Another crucial requirement for truncated SD equations, not implemented yet, is the covariance of  the fermion propagator and the vertex under
the Landau-Khalatnikov-Fradkin (LKF) transformations [\onlinecite{Landau:1955zz,Johnson:1959zz,Sonoda:2000kn}]. These transformations have a simple form
in the coordinate space representation and allow us to evaluate Green's functions in an arbitrary covariant gauge if we
know their value in any particular gauge. However, in momentum space, the LKF transformations have a rather complicated form for
the vertex, and this is the reason why they were not fully involved yet for restricting the form of the vertex  even in the
quenched approximation (for work in this direction, see papers by Bashir and collaborators [\onlinecite{Bashir:2000iq,Bashir:2000rv,Bashir:2002sp}]
and the review in Ref.~[\onlinecite{Bashir:2007zza}]).

In the present paper, we study the gauge-covariance of the massless fermion propagator in quenched QED$_3$ in a linear covariant gauge.
The LKF transformation for unquenched QED$_3$ including a non-local gauge will be considered in the future. At this point, let's recall that the quenched
limit of QED is the approximation in which we can neglect the effects of closed fermion loops. The approximation came
from investigations of the lattice representation of QED$_4$ (see [\onlinecite{Marinari:1981qf,Fucito:1980fh}]) which
showed that a reasonable estimate of the hadron spectrum could be obtained by eliminating all internal quark loops. Moreover, the quenched
approximation in QED$_4$ is now used to include QED effects within lattice QCD calculations (see the recent paper [\onlinecite{Hatton:2020qhk}] and
references and discussions therein). Just after its introduction, the quenched approximation in QED$_4$ has also been used in Refs.~[\onlinecite{Fomin:1984tv,Leung:1985sn,Gusynin1999PRD}] within the formalism of the SD equations.

The paper is organized as follows. In Sec.~\ref{LKF-d=3}, we consider the LKF transformation for the fermion propagator in momentum space in exactly
three dimensions. We notice that it cannot be applied to terms higher than $e^4$ in the perturbative expansion since they lead to infrared singularities
in the LKF relation. Therefore, in Sec.~\ref{sec:QQED-DR}, we apply dimensional regularization (following \cite{Gusynin1999PRD,Kotikov:2019bqo,James:2019ctc})
to deal with higher order terms. The obtained general expressions, valid for arbitrary order terms, are then analyzed thus revealing the self-consistency
of the LKF transformation in Secs.~\ref{sec:LKF1} and \ref{sec:LKF2}. The analysis leads to the conclusion that exactly in three dimensions, $d = 3$, all
odd perturbative coefficients, starting with third order one, should be zero in any gauge. The results are summarized and discussed in Sec.~\ref{Sec:Conclusion}.
In three Appendices \ref{sec:App:A}, \ref{sec:App:B}, and \ref{sec:App:C} we present the details of the calculations.

\section{LKF transformation}
\label{LKF-d=3}

In the following, we shall consider a Euclidean space of dimension $d=3 - 2\veps$.
The general form of the fermion propagator $S_F(p,\xi)$ in some gauge $\xi$ reads:
\be
S_F(p,\xi) = \frac{\I}{\hat{p}} \, P(p,\xi) \, ,
\label{SFp}
\ee
where the tensorial structure, \eg, the factor $\hat{p}$ containing Dirac $\gamma$-matrices, has been extracted and $P(p,\xi)$ is a scalar function
of $p=\sqrt{p^2}$. It is also convenient to introduce the $x$-space representation $S_F(x,\xi)$ of the fermion propagator as:
\be
S_F(x,\xi) =  \hat{x} \, X(x,\xi) \, .
\label{SFx}
\ee
The two representations, $S_F(x,\xi)$ and $S_F(p,\xi)$, are related by the Fourier transform which is defined as:
\begin{subequations}
\label{Fourier:def}
\begin{flalign}
&S_F(p,\xi) =  \int \D^dx \, e^{\I px} \, S_F(x,\xi) \, ,
\label{SFx2p} \\
&S_F(x,\xi) = \int \frac{\D^dp}{(2\pi)^{d} } \, e^{-\I px} \, S_F(p,\xi) \, .
\label{SFp2x}
\end{flalign}
\end{subequations}

In momentum space, the photon propagator can be written in the following general form:
\begin{flalign}
D_{\mu \nu}(q,\xi) = D_{\text{T}}(q^2)\,\left( g^{\mu \nu} - \frac{q^\mu q^\nu}{q^2} \right) + \xi\,D_{\text{L}}(q^2)\,\frac{q^\mu q^\nu}{q^2}\, ,
\label{D:p-space}
\end{flalign}
where the functions $D_{\text{T}}(q^2)$ and $D_{\text{L}}(q^2)$ encode the transverse and longitudinal parts of the photon propagator, respectively,
and the $\xi$-dependence was made explicit. In linear covariant gauges that we shall focus on in this paper,
radiative corrections affect transverse photons and not the longitudinal ones, \ie,
\be
D_{\text{T}}(q^2) = \frac{1}{q^2\,(1 + \Pi(q^2))}\, , \qquad D_{\text{L}}(q^2) = \frac{1}{q^2}\, ,
\label{DT+DL}
\ee
where $\Pi(q^2)$ is the polarization part.

The LKF transformation then expresses the covariance of the fermion propagator under a gauge transformation.
It can be derived by standard arguments, see, \eg, [\onlinecite{Landau:1955zz,Johnson:1959zz,Sonoda:2000kn}]
and its most general form can be written as:
\begin{flalign}
\hspace{-0.2cm} S_F(x,\xi) = S_F(x,\eta)\, G(x,\Delta),\,\, G(x,\Delta) =e^{D(x) - D(0)}\, ,
\label{def:LKF}
\end{flalign}
where
\begin{flalign}
D(x) = e^2 \Delta\, \mu^{2\veps} \int \frac{\D^d q}{(2\pi)^d } \, e^{-\I q x} \, \frac{D_{\text{L}}(q^2)}{q^2}, \,\, \Delta = \xi - \eta\, ,
\label{def:D(x)}
\end{flalign}
thereby relating the fermion propagator in two arbitrary covariant gauges $\xi$ and $\eta$. For $d=3$ and $D_L(q^2)$ given by Eq.~(\ref{DT+DL}),
the LKF transformation takes the form \cite{Bashir:2000iq,Bashir:2007zza}:
\begin{flalign}
S_F(x,\xi) = S_F(x,\eta)\,e^{-a|x|},\quad a=\frac{\alpha\Delta}{2},\quad \alpha=\frac{e^2}{4\pi}\, .
\label{alpha}
\end{flalign}
As can be seen from (\ref{alpha}), the transformation has a very simple form in configuration space.
It is much more complicated in momentum space where the Fourier transform of Eq.~(\ref{def:LKF}) is given by a convolution
\be
S_F(p,\xi)=\int\frac{\D^d k}{(2\pi)^{d}}\,S_F(k,\eta)\,G(p-k,\Delta)\, .
\label{LKF:Fourier}
\ee
Here $G(k,\Delta)$ is the Fourier transform of $G(x,\Delta)$ which, for $d=3$, is given by:
\be
G(k,\Delta)=\frac{8\pi a}{(a^2+k^2)^2}
\label{G-to-delta}
\ee
and is such that $G(k,\Delta=0)=(2\pi)^3\delta^{(3)}(k)$.
In terms of the (scalar) function $P$ of Eq.~(\ref{SFp}), the LKF transform can be written as (still in $d=3$):
\be
P(p,\xi)=\frac{a}{\pi^2}\int \D^3 k\, P(k,\eta)\, \frac{p\cdot k}{k^2[a^2+(p+k)^2]^2}\, .
\label{Pinterd=3}
\ee
Integrating over the angles in Eq.~(\ref{Pinterd=3}) yields a more explicit equation relating the $P$-functions in two different
gauges:
\begin{flalign}
P(p,\xi) &= \frac{a}{\pi}\int^\infty_0 \D k \, P(k,\eta)\, \left[\frac{1}{a^2+(p-k)^2} \right .
\nonum \\
&\left . +\frac{1}{a^2+(p+k)^2}+\frac{1}{2pk}\ln\frac{a^2+(p-k)^2}{a^2+(p+k)^2}\right] \, .
\label{Integ}
\end{flalign}
It is well known that the perturbation theory in $\alpha$ in massless QED$_3$ suffers from infrared
divergences since, in fact, for dimensional reasons, the expansion is in terms of $\alpha/p$ and infrared divergences are encountered as a consequence
of the momenta in the denominator. Potentially infrared divergent diagrams contain insertions of the vacuum polarization diagram
into the gauge propagator. For the fermion propagator, an infrared divergent contribution appears first at the order $\alpha^2$.
However, there are weighty arguments that quenched perturbation theory is infrared finite [\onlinecite{Jackiw1981}]
 {and they have been recently confirmed by lattice studies of the quenched approximation \cite{Karthik2017}.
In the present paper, we therefore focus on the quenched case only.}

Let us use as an initial $\eta$-gauge\,\footnote{Hereafter we call the linear gauge with the parameter
  $\eta$ as the $\eta$-gauge.} the Landau gauge in Eq.~(\ref{Integ}).
At two loop order in quenched perturbation theory,  we have \cite{Bashir:2000iq}:
\be
P(p,0)=1-\frac{3\alpha^2}{4p^2}\left(\frac{7}{3}-\frac{\pi^2}{4}\right),
\ee
where the second term comes from two-loop self-energy diagram with crossed photon lines. Inserting the last expression in
Eq.~(\ref{Integ}) and evaluating the integral we get
\be
P(p,\xi)=1-\frac{a}{p}\arctan\frac{p}{a}-\frac{(28-3\pi^2)\alpha^2p^2}{16(a^2+p^2)^2}.
\label{LKF-from-Landau-gauge}
\ee
Expanding {the} r.h.s.\ up to $O(\alpha^2)$, we reproduce the perturbative expansion for the massless fermion propagator
in an arbitrary covariant gauge
\be
P(p,\xi)=1-\frac{\pi\alpha\xi}{4p}+\frac{\alpha^2\xi^2}{4p^2}+\frac{3\alpha^2}{4p^2}
\left(\frac{\pi^2}{4}-\frac{7}{3}\right)+ {\cal O}(\alpha^4)\, .
\label{B2000}
\ee
Eqs.~(\ref{LKF-from-Landau-gauge}), (\ref{B2000}) are in agreement with Ref.~\cite{Bashir:2000iq}.
Beyond $\alpha^2$, the terms of the $P(p,\xi)$ expansion are not yet calculated in perturbation theory. The LKF transformation, being non-perturbative
in nature, contains, at a given order of the {loop}-expansion, important information about {higher order terms}.
Note, for example, the absence of {the $\alpha^2\xi$ term} and {it has been suggested}
\cite{Bashir:2000iq} {that the contributions $\alpha^3 \xi^m $ ($m=1,2,3$) are also absent upon}
 further expanding the expression (\ref{LKF-from-Landau-gauge}).

Although Eq.~(\ref{Integ}) is an exact one, it cannot be applied to the terms of order $(\alpha/p)^3$ and higher {ones}
in {the loop-}expansion of the massless fermion propagator.
This is because the kernel of Eq.~(\ref{Integ}) (in square brackets) behaves as $\sim k^2$ as $k\to0$ 
 which leads to infrared divergences for higher order terms in the expansion of $P$.
{In the following,} we will show that studying the LKF transformation in dimensional
regularization allows one to get around this difficulty and obtain explicit
finite expressions at any order in quenched perturbation theory for $d=3$. In general, this approach can
also be applied to the unquenched expansion with dimensionally regularized perturbation theory that
we postpone for a future study. 

\section{Quenched QED$_3$ in dimensional regularization}
\label{sec:QQED-DR}

\subsection{LKF exponent $D(x)$}
\label{sec:D(x)}

Following Refs.~\cite{Kotikov:2019bqo} and \cite{James:2019ctc}, we use from now on dimensional regularization for which
$D(0)=0$ ($D(0)$ is a massless tadpole and, thus, it is eliminated in dimensional regularization)
and Eq.~(\ref{def:LKF}) simplifies as:
\be
S_F(x,\xi) = S_F(x,\eta)\,e^{D(x)}\, .
\label{def:LKFd}
\ee
In order to proceed, it is useful to first present the Euclidean space Fourier transforms of massless propagators
(see, for example, the recent review \cite{Kotikov:2018wxe}), which have very simple
and symmetric forms:
%
\begin{subequations}
\label{Fourier}
\begin{flalign}
&\int \frac{\D^d q}{(2\pi)^d } \, \frac{e^{-\I q x}}{(q^2)^{\beta}} =
\frac{1}{(4\pi)^{d/2}} \, \frac{2^{2\tilde \beta} \, a(\beta)}{(x^2)^{\tilde \beta}} \, ,
\label{Fourier1} \\
&\int \frac{\D^d x}{(2\pi)^d } \, \frac{e^{\I q x}}{(x^2)^{\beta}} =
\frac{1}{(4\pi)^{d/2}} \, \frac{2^{2\tilde \beta}a(\beta)}{(q^2)^{\tilde \beta}} \, ,
\label{Fourier2}
\end{flalign}
\end{subequations}
where
\be
a_n(\beta)=\frac{\Gamma(\tilde \beta+n)}{\Gamma( \beta)},~~a(\beta)= a_0(\beta),
~~\tilde \beta =\frac{d}{2} -\beta \, .
\label{tbeta}
\ee
%
With the help of (\ref{Fourier1}), we may evaluate $D(x)$ which is expressed as
\be
D(x) = e^2\,\Delta\,\mu^{2\veps}\,\int \frac{\D^d q}{(2\pi)^d } \, \frac{e^{-\I q x}}{q^4} \, ,
\label{def:D(x):QED3}
\ee
in $d=3-2\ep$. The calculation yields \cite{Gusynin1999PRD}:
%
\begin{flalign}
&D(x) = \frac{\al\,\Delta}{4\pi}\,\mu^{2\ep}\,\Gamma(d/2-2)\,(\pi\,x^2)^{2-d/2}
\nonum \\
&= - K(\pi \mu^2 x^2)^{1/2+\ep},\,K=\frac{\al\,\Delta}{2\pi\,\mu}\,\frac{\Gamma(1/2-\ep)}{1+2\ep} \, ,
\label{DxQED3}
\end{flalign}
which, for $\ep=0$, leads to the exponent in Eq.~(11).

Notice that Eq.~(\ref{DxQED3}) is finite in the limit $\ep \ra 0$, \ie, QED$_3$ is free from UV singularities.
Moreover, in the quenched case that we consider, IR singularities arising from fermion loops, \eg,
in gauge-invariant contributions that are in the pre-exponential factor of (\ref{def:LKFd}), are suppressed.
Nevertheless, as anticipated in the Introduction, the super-renormalizable nature of quenched QED$_3$ may still
give rise to IR singularities at high orders. In dimensional regularization, these IR singularities
take the form of poles in $1/\ep$ just as UV singularities. In principle, in order to keep track of them, the
full $\ep$-dependence of Eq.~(\ref{DxQED3}) has to be taken into account.

\subsection{Momentum space LKF transformation}
\label{sec:p-space-LKF}

Following the usual steps for the derivation of the LKF transformation,
we then consider the fermion propagator $S_F(p,\eta)$ with the external momentum $p$ in some gauge $\eta$.
The latter has the form (\ref{SFp}) with $P(p,\eta)$ as
\be
P(p,\eta) = \sum_{m=0}^{\infty} a_m(\eta)\, \left(\frac{\al}{2\sqrt{\pi}\,p}\right)^m\, {\left(\frac{\tilde{\mu}^2}{p^2}\right)}^{m\ep} \, ,
\label{Peta:QED3}
\ee
where $a_m(\eta)$ are coefficients of the loop expansion of the propagator and
$\tilde{\mu}$ is the scale
%
\be
\tilde{\mu}^2= 4\pi \mu^2 \, ,
\label{Aem}
\ee
which is intermediate between the MS scale $\mu$ and the $\overline{\rm{MS}}$ scale $\overline{\mu}$.
In Eq.~(\ref{Peta:QED3}), the expansion has been
written in terms of the dimensionless ratio $\al / p$ with an additional conventional factor of $1/(2\sqrt{\pi})$.
Its exact form is coming from the consideration of four-dimensional QED in \cite{Kotikov:2019bqo}
(see also App.~\ref{sec:App:A} and discussions just after Eq.~(\ref{DxQEDd})).

  Now we would like to find the exact formulas for the transformation  $a_m(\eta) \to a_m(\xi)$. We will do it in two
  different ways. Following Refs.~\cite{Kotikov:2019bqo,James:2019ctc},
  in the following first
  sub-subsection
  we will obtain it from the $x$-space LKF evolution (\ref{def:LKFd}) transforming the ansatz (\ref{Peta:QED3}) to $x$-space.
  Since only few details of the approach were given 
  in \cite{Kotikov:2019bqo}, 
 the full demonstration will be given in App.~\ref{sec:App:A} in the case of an Euclidean space of arbitrary dimension $d$.
  The second sub-subsection, together with App.~\ref{sec:App:B} for a more detailed analysis, will provide the corresponding
  derivation in momentum-space with the help of a direct evaluation of $G(p,\Delta)$
  and then the evaluation Eq.~(\ref{LKF:Fourier}).

\subsubsection{$x$-space analysis}
\label{sec:LKF:x-space-analysis}

Using the ansatz (\ref{Peta:QED3}) and the Fourier transform (\ref{Fourier3}),
we obtain that (see App.~\ref{sec:App:A} for a more extended analysis in arbitrary $d$):
\begin{flalign}
S_F(x,\eta) = \frac{2^{d-1}\,\hat{x}}{(4\pi\, x^2)^{d/2}} \,
\sum_{m=0}^{\infty} b_m(\eta)\, \, \left(\frac{\al\,x}{4\sqrt{\pi}}\right)^m\,
{\left(\pi\mu^2x^2\right)}^{m\ep} \, ,~~
\label{SFetad3}
\end{flalign}
with
\be
b_m(\eta) =  a_m(\eta) \, \frac{\Gamma(3/2-m/2-(m+1)\ep)}{\Gamma(1+m/2+ m\ep)} \, .
\label{bm.etad3}
\ee

Expanding the LKF exponent, we have
\begin{flalign}
&S_F(x,\xi)=
S_F(x,\eta) e^{D(x)} = \frac{2^{d-1}\,\hat{x}}{(4\pi\, x^2)^{d/2}} \,
\nonumber \\
&\times
\sum_{m=0}^{\infty} b_m(\eta)\, \, \left(\frac{\al\,x}{4\sqrt{\pi}}\right)^m\,
    {\left(\pi\mu^2x^2\right)}^{m\ep} \,\nonumber \\
&\times
    \sum_{l=0}^{\infty}
    {\left(- \frac{\al \,\Delta x}{2\sqrt{\pi}}\right)}^l\, \phi(l,\ep) \,
    (\pi \mu^2 x^2)^{l\ep}
\, ,
\label{SFxid3.0}
\end{flalign}
where
\be
\phi(l,\ep) = \frac{\Gamma^l(1/2-\ep)}{l!\,(1+2\ep)^l}\, .
\label{phi}
\ee
Factorizing all the $x$-dependence yields:
\begin{flalign}
S_F(x,\xi) = \frac{2^{d-1}\,\hat{x}}{(4\pi\, x^2)^{d/2}} \,
\sum_{k=0}^{\infty} b_k(\xi)\,  \left(\frac{\al\,x}{4\sqrt{\pi}}\right)^k \, {\left(\pi\mu^2x^2\right)}^{k\ep} \, ,~~
\label{SFxid3}
\end{flalign}
with
\be
b_k(\xi) =  \sum_{m=0}^{k} \, (-2 \Delta)^{k-m} \, b_m(\eta) \,   \phi(k-m,\ep)
\, .
\label{bm.xid3}
\ee
Hence, from the correspondence between the results for the propagators
$P(p,\eta)$ and $S_F(x,\eta)$ in (\ref{Peta:QED3}) and
(\ref{SFetad3}), respectively, together with the result (\ref{SFxid3}) for $S_F(x,\eta)$,
we have, for $P(p,\xi)$:
\be
P(p,\xi) = \sum_{k=0}^{\infty} a_k(\xi)\,  \left(\frac{\al}{2\sqrt{\pi}\,p}\right)^k \,
{\left(\frac{\tilde{\mu}^2}{p^2}\right)}^{k\ep} \, ,
\label{Pxid3}
\ee
where
\begin{flalign}
&a_k(\xi) = b_k(\xi) \, \frac{\Gamma(1+k/2+k\ep)}{\Gamma(3/2-k/2-k\ep)}
\nonum \\
&=  \sum_{m=0}^{k}
\, (-2 \Delta)^{k-m} \, a_m(\eta) \, \hat{\Phi}(m,k,\ep) \  \phi(k-m,\ep)
\label{am.xid3}
\end{flalign}
and
\begin{flalign}
\hat{\Phi}(m,k,\ep)
= \frac{\Gamma(3/2-m/2-(m+1)\ep)\Gamma(1+k/2+k\ep)}{\Gamma(1+m/2+m\ep)
  \Gamma(3/2-k/2-(k+1)\ep)} \, .
\label{hPhi}
\end{flalign}
In this way, we have derived the expression of $a_k(\xi)$ using a simple expansion of the LKF exponent in $x$-space.
From this representation of the LKF transformation, we see that the magnitude $a_k(\xi)$ is determined by $a_m(\eta)$ with $0 < m<k$.

We would like to note that Eq.~(\ref{Pxid3}) exactly reproduces our initial ansatz (\ref{Peta:QED3}). It shows
that the initial ansatz (\ref{Peta:QED3}) is correct. Moreover, the representation (\ref{Pxid3}) can be used as starting
point for the evolution to another gauge (this will be discussed in Sec.~\ref{sec:LKF2}).

Very often, however (see Eqs.~(\ref{LKF-from-Landau-gauge}) and (\ref{B2000})
and Ref.~\cite{Bashir:2000iq}), the subject of the study is not the magnitude $a_m(\xi)$ but the $p$- and $\Delta$-dependencies of each
magnitude $a_l(\eta)$ as it evolves from the $\eta$ to the $\xi$ gauge. The corresponding result for the  $p$- and $\Delta$-dependencies
of $\hat{a}_m(\xi,p)$ can be obtained interchanging the order of the sums in the r.h.s.\ of (\ref{Pxid3}). Performing such interchange yields:
\be
P(p,\xi) = \sum_{m=0}^{\infty} \hat{a}_m(\xi,p)\, \left(\frac{\al}{2\sqrt{\pi}\,p}\right)^m  \,
{\left(\frac{\tilde{\mu}^2}{p^2}\right)}^{m\ep} \, ,
\label{Pxi.1d3}
\ee
where, now, the coefficients transform as
\begin{flalign}
\hat{a}_m(\xi,p) = a_m(\eta) \,\sum_{l=0}^{\infty}\tilde{\Phi}(m,l,\ep) \, \phi(l,\ep) \,
\left(-\frac{\alpha \Delta}{\sqrt{\pi}p}\right)^l \, {\left(\frac{\tilde{\mu}^2}{p^2}\right)}^{l\ep}\, ,
\label{axi:QED3}
\end{flalign}
with
%
\begin{flalign}
&
\tilde{\Phi}(m,l,\ep) = \hat{\Phi}(m,m+l,\ep)
\label{tPhi} \\
&\hspace{-0.5cm}=\frac{\Gamma(3/2-m/2-(m+1)\ep)\Gamma(1+(m+l)/2+(m+l)\ep)}{\Gamma(1+m/2+m\ep)\Gamma(3/2-(m+l)/2-(m+l+1)\ep)} \, .
\nonum
\end{flalign}
%

%

\subsubsection{$p$-space analysis}

Here we present the basic steps of the direct derivation of Eq.~(\ref{axi:QED3}). A more extended analysis
can be found in Appendix B.

To evaluate the Fourier transform of $e^{D(x)}$ we use the Mellin integral representation,
\bea
&&e^{-b (x^2)^{\nu}}=\frac{1}{2\pi \I}\hspace{-2mm}\int\limits_{\gamma-\I \infty}^{\gamma+\I \infty}\hspace{-2mm}\D s\, b^{-s}\frac{\Gamma(s)}{x^{2\nu s}}\, ,
\quad{\rm Re}\,\gamma>0,\nonumber\\
&& b=K(\pi\mu^2)^\nu>0,\quad \nu=1/2+\epsilon>0\, ,
\label{Mellin-int-D}
\eea
integrate over $x$ and get (see App.~\ref{sec:App:B} for the details):
\begin{flalign}
G(p,\Delta) = \left(\mu^2K^{\frac{1}{\nu}}\right)^{-\frac{d}{2}}\,
\mathcal{E}_{\nu,\frac{d}{2}}\left( - \frac{p^2}{\tilde{\mu}^2\,K^{\frac{1}{\nu}}} \right)\, .
\label{G(p):exact}
\end{flalign}
The two-parameter function ${\cal E}_{\nu,\alpha}(z)$ is a special case of the generalized Wright function ${}_1\Psi_1$ \cite{Kilbas2002}.
Since the LKF transformation is non-perturbative in nature, the exact expression (\ref{G(p):exact}) can be used when solving, for example,
truncated Schwinger-Dyson equations.

To evaluate the convolution integral Eq.~(\ref{LKF:Fourier}) we use the Mellin-Barnes representation of this function
\be
{\cal E}_{\nu,\alpha}(-z) =
\frac{1}{2\pi \I}\int\limits_{\gamma - \I \infty}^{\gamma + \I \infty} \D s \,z^{-s}\,
\frac{\Gamma(s)\Gamma\left(\frac{\alpha-s}{\nu}\right)}{\nu\,\Gamma(\alpha-s)}\, ,
\label{MB-Efunction}
\ee
where the contour separates the poles of the gamma functions in the numerator.
By using the expansion (\ref{Peta:QED3}), together with Eqs.~(\ref{G(p):exact}) and (\ref{MB-Efunction}),
we can integrate over $k$ and get the final answer for $P(p,\xi)$ in the form of (\ref{Pxi.1d3}),
%
with the
coefficients $\hat{a}_m(\xi,p)$ representing the gauge-evolution of the magnitude $a_m(\eta)$:
\bea
\hat{a}_m(\xi,p)&=&a_m(\eta)\frac{\Gamma(3/2-\epsilon-m\nu)}{\nu\Gamma(1+m\nu)}\left(\frac{p^2}{\tilde{\mu}^2\,K^{\frac{1}{\nu}}}\right)^{1+m\nu}\nonumber\\
&\times&{}_1\Psi_1\left[-\frac{p^2}{\tilde{\mu}^2\,K^{\frac{1}{\nu}}}\Bigr|\begin{array}{c}(m+\frac{1}{\nu},\frac{1}{\nu})\\ (5/2-\epsilon,1)
\end{array}\right]\, .
\label{a_m-relation_final_explicit}
\eea
Using the asymptotic expansion of the Wright function at large values of its argument,
we can write the LKF relation between
the coefficients $\hat{a}_m(\xi,p)$ and $a_m(\eta)$ in the form (\ref{axi:QED3}), (\ref{tPhi}).

\section{Analysis of
  Eq.~(\ref{Pxi.1d3})}
\label{sec:LKF1}

In this section we analyze the LKF transformation represented by Eq.~(\ref{Pxi.1d3}) together with
(\ref{axi:QED3}) and (\ref{tPhi}).

\subsection{Coefficients $\hat{a}_k(\xi,p)$ at $\ep=0$}
\label{Sec:ha-ep=0}

As discussed in Sec.~\ref{Sec:Introduction}, quenched QED$_3$ is {\it a priori} free from both IR and UV singularities.
Because of this finiteness, it is tempting to set $\ep=0$ in Eq.~(\ref{a_m-relation_final_explicit}).
We consider this possibility here as a prerequisite to the more complete analysis of the following (sub-)sections.

In the case $\ep=0$, the fermion propagator (\ref{Pxi.1d3}) takes the simpler form
\be
P(p,\xi) = \sum_{m=0}^{\infty} \hat{a}_m(\xi,p) \, \left(\frac{\al}{2\sqrt{\pi}\,p}\right)^m \, ,
\label{Pxi:ha-ep=0}
\ee
where the coefficients (\ref{a_m-relation_final_explicit}) are greatly simplified since the function ${}_1\Psi_1$ reduces to the Gauss
hypergeometric function. Hence, we get:
\begin{flalign}
&\hat{a}_m(\xi,p) = a_m(\eta)\,B^{-(2+m)}\,\frac{1 - m^2}{3\,\cos (\pi\,m/2)}\,
\nonum \\
&\times\,{}_2 F_1 \left( \frac{m+2}{2}, \frac{m+3}{2}; \frac{5}{2}; - \frac{1}{B^2} \right),\, B=\frac{\alpha\Delta}{2p}\, .
\label{axi.ep=0-tot2}
\end{flalign}
The last expression can be rewritten in a slightly different form using analytic
continuation through connection formulas for ${}_2 F_1$ hypergeometric functions (see, \eg, \cite{Bateman1})
:\,\footnote{
  We would like to note that these equations can also be obtained by considering even and odd $l$-values
in Eq.~(\ref{axi:QED3}) at $\ep =0$.}
\begin{flalign}
\hat{a}_m(\xi,p) &= a_m(\eta)\,\Biggr[{}_2 F_1 \left( \frac{m-1}{2}, \frac{m+2}{2}, \frac{1}{2}; - B^2 \right)
 \nonum \\
&\hspace{-1.4cm}+ B\,\frac{m^2 - 1}{m}\,\tan \left(\frac{\pi m}{2} \right)\,
{}_2 F_1 \left( \frac{m}{2}, \frac{m+3}{2}; \frac{3}{2}; - B^2 \right) \Biggl]\, .
\label{axi.ep=0-tot}
\end{flalign}
Eqs.~(\ref{axi.ep=0-tot2}) and (\ref{axi.ep=0-tot}) are suitable for describing series expansions at  small
and large values of $p$, respectively.

Let's end this subsection by giving some explicit formulas for the first values of $m$ from Eq.~(\ref{axi.ep=0-tot2})
(see, \eg, \cite{Prudnikov3} for the evaluation of ${}_2F_1$ hypergeometric functions):
\begin{subequations}
\label{axi.QuenchedQED3.2-explicit}
\begin{flalign}
\hat{a}_0(\xi,p) &= a_0(\eta)\,\biggl( 1 - B\,\arctan ( 1/B ) \biggr)
\nonum \\
&= a_0(\eta)\,\biggl( 1 - \frac{\pi B}{2} +  B^2 + \Ord(B^3) \biggr)\, ,
 \\
\hat{a}_1(\xi,p) &= -a_1(\eta)\,\frac{2}{\pi}\biggl( \frac{B}{(1 + B^2)}- \arctan ( 1/B ) \biggr)
\nonum \\
&= a_1(\eta)\,\biggl( 1 - \frac{4 B}{\pi} + \Ord(B^3) \biggr)\, ,
 \\
\hat{a}_2(\xi,p) &= a_2(\eta)\,\frac{1}{(1 + B^2)^2}
\nonum \\
&= a_2(\eta)\,\biggl( 1 - 2B^2 + \Ord(B^3) \biggr)\, ,
\\
\hat{a}_4(\xi,p) &= a_4(\eta)\frac{1-5B^2}{(1+B^2)^4}
\nonum \\
&=\hat{a}_4(\eta)\,\biggl(1-9B^2+\Ord(B^4)\biggr)\, .
\end{flalign}
\end{subequations}
The results (\ref{axi.QuenchedQED3.2-explicit}) are in exact agreement with those derived from
the integral relation (\ref{Integ}). Thus, substituting back (\ref{axi.QuenchedQED3.2-explicit}) in the fermion
propagator (\ref{Pxi:ha-ep=0}), the LKF transformation is found to be in perfect agreement with the perturbative result
(\ref{B2000}).

Starting from $m \geq 3$, both Eqs.~(\ref{axi.ep=0-tot}) and (\ref{axi.ep=0-tot2}) are subject to singularities. This agrees with the analysis based
on (\ref{Integ}) which shows that such singularities are of IR nature and arise
from the super-renormalizability of (quenched) QED$_3$. Moreover, from Eq.~(\ref{axi.ep=0-tot2}) --
and equivalently from (\ref{axi.ep=0-tot})
-- a parity effect is clearly displayed whereby even coefficients, $\hat{a}_{2s}(\xi,p)$, are finite for all $s$
while odd coefficients, $\hat{a}_{2s+1}(\xi,p)$, are {\it singular} for all $s \geq 1$.

Such an observation calls for a more careful treatment of quenched QED$_3$. In particular, the parameter $\ep$
has to be kept non-zero in order to regulate the singularities. This will be the subject of the following
subsection.

\subsection{Coefficients $\hat{a}_{m}(\xi,p)$ at $\ep \ra 0$}
\label{Sec:ha-ep->0}


We would like to complete the results of the previous subsection by computing the coefficients $\hat{a}_{m}(\xi,p)$ at $\ep \ra 0$.
A subtlety in the related $\ep$-expansion is that only the leading order term in $\ep \ra 0$ (whether a constant, a pole $1/\ep$ and/or a
  contribution $\sim \ep$) is needed in order to analyze the previously described parity effect
(between odd and even coefficients $\hat{a}_{m}(\xi,p)$). Such an expansion strategy will be fully appreciated (and detailed)
in the next subsection.



Interestingly, from Eq.~(\ref{a_m-relation_final_explicit}),
we see quite clearly that singularities do affect odd coefficients, $m=2s+1$ for $s\geq 1$, and that these
singularities are entirely located in the first gamma function in the numerator.
Other terms including the Wright function, ${}_1\Psi_1$, are finite.
Postponing the details of the derivations to App.~\ref{sec:App:B},
in the following we shall be brief and present only the final results.

The finite even coefficients, $\hat{a}_{2s}$, may be re-written for $\ep=0$ as (see, Eq.~(\ref{axi.ep=0-tot})):
\bea
&&\hat{a}_{2s}(\xi,p) = a_0(\eta)\,\delta_0^s\,\left[1-B\arctan(1/B)\right]\nonumber\\
&&+(1-\delta_0^s)a_{2s}(\eta) {}_2F_1\left(s-\frac{1}{2},s+1;\frac{1}{2};-B^2\right).
\label{ha2s:even}
\eea
where $\delta_0^s$ is the Kronecker symbol.

In the case of the odd coefficients, $m=2s+1$ ($s\geq 1$), the leading order result takes the form of a simple pole in
$\ep$. It follows from Eq.~(\ref{a_m-relation_final_explicit}) that:
\bea
&&\hat{a}_{2s+1}(\xi,p)
= \frac{2s(-1)^s}{3\pi\ep} \, a_{2s+1}(\eta)\,B^{-(3+2s)} \nonumber\\
&&\times {}_2F_1\left(s+2,s+\frac{3}{2};\frac{5}{2}; -\frac{1}{B^2}\right).
\label{ha2s1:odd}
\eea
This can be re-written in slightly different form as:
\bea
&&\hat{a}_{2s+1}(\xi,p)
= - \frac{2sB}{(2s+1)\pi\ep} \, a_{2s+1}(\eta)\nonumber\\
&&\times {}_2F_1\left(s+2,s+\frac{1}{2};\frac{3}{2}; -B^2\right)\, .
\label{F21:1c}
\eea

In agreement with the previous subsection, the above results confirm that odd coefficients $\hat{a}_{2s+1}(\xi,p)$ are singular for $s\geq 1$, see
(\ref{ha2s1:odd}), while even coefficients are finite. Assuming that quenched QED$_3$ is finite
(see Refs~.[\onlinecite{Jackiw1981}] and discussion therein), we find that the requirement for the limit
$\ep=0$ to be well-defined is:
\be
a_{2s+1}(\xi)=0 \qquad \forall s \geq 1\, ,
\label{res:odda=0}
\ee
\ie, the identical vanishing of all odd coefficients (magnitudes) other than $a_{1}(\xi)$.
In the next section, we will provide a refined proof of  (\ref{res:odda=0}) based on the direct
analysis of the coefficients $a_k(\xi)$ (rather than their gauge evolution $\hat{a}_{m}(\xi,p)$).

\section{Analysis of
  Eq.~(\ref{Pxid3})}
\label{sec:LKF2}

In this section we analyze the LKF transformation represented by Eq.~(\ref{Pxid3}) together
with Eqs.~(\ref{am.xid3}) and (\ref{hPhi}).

\subsection{Coefficients $a_k(\xi)$ at $\ep \ra 0$}
\label{sec:LKF2a}

The analysis of the parity effect may be pushed further with the help of the first
representation of the LKF transform,
Eq.~(\ref{Pxid3}), that allows to study the momentum-independent magnitudes, $a_k(\xi)$, of Eq.~(\ref{am.xid3}).
As for the study of their gauge-evolution, $\hat{a}_{m}(\xi,p)$, the $\ep$-expansion of $a_k(\xi)$ will be carried
out at leading order in $\ep \ra 0$. In the following, we shall justify such a procedure by showing that this
accuracy is enough to prove the self-consistency of the LKF transformation.\\

The analysis of the coefficients $a_k(\xi)$ requires considering the cases of even and odd values of $k$ separately.
The complete evaluation is carried out in App.~\ref{sec:App:C}. Here we present only the final results.\\

{
{\bf 1.}~ In the case of even $k$ values, \ie, $k=2r$,
the final results for $a_{2r}(\xi)$
can be expressed as a sum of the contributions
  $a_{2r}^{(i)}(\xi)$ with $i=1,2$ and $3$, \ie,
\be
a_{2r}(\xi) = a_{2r}^{(1)}(\xi) + a_{2r}^{(2)}(\xi) + a_{2r}^{(3)}(\xi)\, .
\label{axi:a2s0}
\ee
The latter come in-turn from the corresponding contributions of the initial
amplitudes $a_{2s}(\eta)$, $a_{1}(\eta)$ and $a_{2s+1}(\eta)$ as}
\begin{subequations}
\label{axi:a2s}
\begin{flalign}
&a_{2r}^{(1)}(\xi) = \sum_{s=0}^{r} \, a_{2s}(\eta) \,
\nonum \\
  &\times\, \frac{\Gamma(r-1/2)\Gamma(1+r)}{\Gamma(1+s))\Gamma(s-1/2)} \,
\frac{\Gamma(1/2)\, (-\delta^2)^{r-s}}{\Gamma(r-s+1/2)(r-s)!} \, ;
\label{axi:a2s1} \\
&a_{2r}^{(2)}(\xi) = \frac{2}{\pi} \, \frac{r}{r-1/2} \, \frac{(-\delta^2)^{r}}{\delta} \, a_{1}(\eta) \, ;
\label{axi:a2sa1} \\
&a_{2r}^{(3)}(\xi) = \sum_{s=1}^{r-1} \, a_{2s+1}(\eta) \,  \frac{(-1)^{r+s+1}}{2(s+1) \pi \ep} \,
\nonum \\
&\times\,\frac{\Gamma(r-1/2)\Gamma(1+r)}{\Gamma(s)\Gamma(s+3/2)} \,
\frac{\Gamma(1/2)\, (-\delta)^{2r-2s-1}}{\Gamma(r-s)\Gamma(r-s+1/2)}
\, ,
\label{axi:a2s3}
\end{flalign}
\end{subequations}
where $\delta = \sqrt{\pi} \Delta$.\\
%

{
 {\bf 2.}~  In the case of odd $k$ values, \ie, $k=2r+1$, we should consider the cases $r=0$ and  $r \geq 1$ separately.}

{
In the case $k=1$, we have the following result:}
\begin{flalign}
a_1^{(1)}(\xi) =  a_1(\eta) \, - \frac{\pi}{2} \, \delta \, a_0(\eta)\, .
\label{axi:1f}
\end{flalign}

{
The final result for $a_{2r+1}(\xi)$ (for $r \geq 1$)
can be expressed as a sum of the contributions
$a_{2r+1}^{(i)}(\xi)$ with $i=1$, $2$ and $3$, \ie,
\be
a_{2r+1}(\xi) = a_{2r+1}^{(1)}(\xi) + a_{2r+1}^{(2)}(\xi) + a_{2r+1}^{(3)}(\xi)\, .
\label{axi:a2s+1}
\ee
The latter come in-turn from the corresponding contributions of the initial
amplitudes $a_{2s}(\eta)$, $a_{1}(\eta)$ and $a_{2s+1}(\eta)$ as}
\begin{subequations}
\label{axi:a2s+11}
\begin{flalign}
&a_{2r+1}^{(1)}(\xi) = [2\pi(r+1)\ep] \,
\sum_{s=0}^{r} \, a_{2s}(\eta) \, (-1)^{r+s+1}
\nonum \\
&\times\,\frac{\Gamma(r+3/2)\Gamma(r)}{\Gamma(1+s))\Gamma(s-1/2)} \,
\frac{\Gamma(1/2)\, (-\delta)^{2r-2s+1}}{\Gamma(r-s+1)\Gamma(r-s+3/2)}\, ;
\label{axi:a2s11} \\
&a_{2r+1}^{(2)}(\xi) = [4(r+1)\ep] \, \frac{r+1/2}{r} \, (-\delta^2)^r \, a_1(\eta)\, ; \label{axi:a2s12}\\
& a_{2r+1}^{(3)}(\xi) = \sum_{s=1}^{r} \, a_{2s+1}(\eta) \, \frac{(r+1)}{(s+1)} \,
\frac{\Gamma(r+3/2)\Gamma(r)}{\Gamma(s)\Gamma(s+3/2)}
\nonum \\
&\times\,\frac{\Gamma(1/2)\, (-\delta^2)^{r-s}}{\Gamma(r-s+1)\Gamma(r-s+1/2)}\, .
\label{axi:a2s13}
\end{flalign}
\end{subequations}
{
We note that, as anticipated above, the contributions (\ref{axi:a2s}), (\ref{axi:1f}) and (\ref{axi:a2s+11})
 {correspond to the first terms of the $\ep$-expansion},
  which is sufficient to analyze the self-consistency given
in the next subsection.}\\

\subsection{Self-consistency}

Consider $a_{m}(\xi)$ with $m\leq 6$. Using the results of the previous subsection, we have:
%
\begin{subequations}
\label{axi16}
\begin{flalign}
&a_{0}(\xi)=a_{0}(\eta)\, ,
\label{sc:axi0} \\
&a_{1}(\xi)=a_{1}(\eta) - \frac{\pi}{2} \, \delta \, a_{0}(\eta)\, ,
\label{sc:axi1} \\
&a_{2}(\xi)=a_{2}(\eta) - \frac{4}{\pi} \, \delta \, a_{1}(\eta) + \delta^2 \, a_{0}(\eta)\, ,
\label{sc:axi2} \\
&a_{3}(\xi)=a_{3}(\eta)  + 6\pi\ep \, \delta \, a_{2}(\eta) - 12\ep \, \delta^2 \, a_{1}(\eta)
\nonum \\
&\qquad ~~~\,+2\pi\ep \, \delta^3 \, a_{0}(\eta)\, ,
\label{sc:axi3} \\
&a_{4}(\xi)=a_{4}(\eta) - \frac{2\delta}{3\pi\ep} \, a_{3}(\eta) - 2 \delta^2 \, a_{2}(\eta)
+ \frac{8\delta^3 }{3\pi} \, a_{1}(\eta)
\nonum \\
&\qquad ~~~\,-\frac{\delta^4 }{3} \, a_{0}(\eta)\, ,
\label{sc:axi4} \\
&a_{5}(\xi)=a_{5}(\eta)  + \frac{45}{2}\pi\ep \, \delta \, a_{4}(\eta) - \frac{15}{2} \, \delta^2 \, a_{3}(\eta)
\nonum \\
&\qquad ~~~\,-15\pi \ep \, \delta^3 \, a_{2}(\eta) + 15\ep \, \delta^4 \, a_{1}(\eta)
- \frac{3}{2} \pi\ep \, \delta^5 \, a_{0}(\eta)\, ,
\label{sc:axi5} \\
&a_{6}(\xi)=a_{6}(\eta) + \frac{4\delta}{5\pi\ep} \, a_{5}(\eta) - 9 \delta^2 \, a_{4}(\eta)
+ \frac{2\delta^3 }{\pi\ep} \, a_{3}(\eta)
\nonum \\
&\qquad ~~~\,+3 \delta^4 \, a_{2}(\eta)
- \frac{12\delta^5 }{5\pi} \, a_{1}(\eta) + \frac{\delta^6 }{5} \, a_{0}(\eta)\, .
\label{sc:axi6}
\end{flalign}
\end{subequations}
Remarkably, these equations are self-consistent. For example, if we would like to obtain the expression
of $a_{m}(\xi_1)$ in some gauge with parameter $\xi_1$ (\ie, the $\xi_1$-gauge), we can derive
it from the $\eta$-gauge using Eq.~(\ref{axi16}) with the replacement $\xi \to \xi_1$ and then proceed in two steps:
from the $\eta$-gauge to the $\xi$-gauge (with help of Eq.~(\ref{axi16})) and later from the $\xi$-gauge to the $\xi_1$-gauge
(with help of Eq.~(\ref{axi16}) with the replacements $\xi \to \xi_1$ and $\eta \to \xi$).

Let's show explicitly this self-consistency in the case of $a_{m}(\xi_1)$ with $m=0,1,2$.
The coefficient $a_{0}(\xi_1)$ does not change, \ie,
\be
a_{0}(\xi_1)=a_{0}(\xi)=a_{0}(\eta)\, .
\label{axi10}
\ee
For the coefficient $a_{1}(\xi_1)$, we have (hereafter $\overline{\delta}_1=\sqrt{\pi}(\xi_1-\xi)$,
$\delta_1=\sqrt{\pi}(\xi_1-\eta)$):
\begin{flalign}
a_{1}(\xi_1) &= a_{1}(\xi)-\frac{\pi}{2} \, \overline{\delta}_1 \, a_{0}(\xi)
\nonum \\
&=
\Bigl(a_{1}(\eta) - \frac{\pi}{2} \, \delta \, a_{0}(\eta)\Bigr) -\frac{\pi}{2} \, \overline{\delta}_1 \, a_{0}(\eta)
\nonum \\
&= a_{1}(\eta) - \frac{\pi}{2} \, \delta_1 \, a_{0}(\eta) \, ,
\label{axi11}
\end{flalign}
because
\be
\delta_1 = \delta + \overline{\delta}_1\, .
\label{delta1}
\ee
So, we obtain the expression of $a_{1}(\xi_1)$ and it coincides with the one obtained directly from the $\eta$-gauge with
help of Eq.~(\ref{sc:axi1}) (with the replacements $\xi \to \xi_1$ and $\eta \to \xi$).

Similarly, the coefficient  $a_{2}(\xi_1)$ changes as:
\begin{flalign}
a_{2}(\xi_1) &= a_{2}(\xi)-\frac{4}{\pi} \, \overline{\delta}_1 \, a_{1}(\xi) + \overline{\delta}^2_1 \, a_{1}(\xi)
\nonum \\
&= \Bigl(a_{2}(\eta) - \frac{4}{\pi} \, \delta \, a_{1}(\eta) + \delta^2 \, a_{0}(\eta)\Bigr)
\nonum \\
&\qquad -\frac{4}{\pi} \,
\Bigl(a_{1}(\eta) - \frac{\pi}{2} \, \delta \, a_{0}(\eta)\Bigr) + \overline{\delta}^2_1 \, a_{1}(\eta) \, .
\label{axi12}
\end{flalign}
The term in factor of $a_{1}(\eta)$ corresponds to:
\be
-\frac{4}{\pi} \, \overline{\delta}_1 -\frac{4}{\pi} \, \delta = -\frac{4}{\pi} \, \delta_1 \, ,
\label{axi11.1}
\ee
because of (\ref{delta1}).
The term in factor of $a_{2}(\eta)$ corresponds to:
\be
\overline{\delta}^2_1 + 2 \overline{\delta}_1\delta + \delta^2 = (\overline{\delta}_1 + \delta)^2 = \delta_1^2
\, .
\label{axi12.1}
\ee
Taking all the results together, we have:
\be
a_{2}(\xi_1)= a_{2}(\eta) - \frac{4}{\pi} \, \delta_1 \, a_{1}(\eta) + \delta_1^2 \, a_{0}(\eta)\, .
\label{axi12.tot}
\ee
Thus, we derive the expression of $a_{2}(\xi_1)$ and it coincides with the one obtained directly from the $\eta$-gauge with
help of Eq.~(\ref{sc:axi2}) (with the replacements $\xi \to \xi_1$ and $\eta \to \xi$).

Similar transformations can also be performed for the other coefficients $a_{i}(\xi_1)$ $(i\geq 2)$ in a similar way.
So, we can obtain a full agreement between the transformation and the results for $a_{i}(\xi_1)$ obtained directly from the $\eta$-gauge with
help of Eqs.~(\ref{axi16}) (with the replacements $\xi \to \xi_1$ and $\eta \to \xi$).

A central result to the present study that can be derived from Eqs.~(\ref{axi16}) is that,
excepting the case of $a_{1}(\xi)$, all $a_{2m+1}(\xi)$ can be excluded. Indeed,
{\it assuming that quenched QED is both UV and IR finite},
(see, for example, Refs.~[\onlinecite{Jackiw1981}] and discussion therein)
{\it setting $\ep=0$ enforces $a_{2m+1}(\xi)=0$ for $(m\geq 1)$}, \ie, (\ref{res:odda=0}) is realized. It follows then that simpler expressions
are obtained for $a_{2i}(\xi)$ with $(i\geq 2)$:
\begin{subequations}
\begin{flalign}
&a_{4}(\xi)=a_{4}(\eta)
- 2 \delta^2 \, a_{2}(\eta)
+ \frac{8\delta^3 }{3\pi} \, a_{1}(\eta) -  \frac{\delta^4 }{3} \, a_{0}(\eta)\, ,
 \\
&a_{6}(\xi)=a_{6}(\eta)
- 9 \delta^2 \, a_{4}(\eta)
+ 3 \delta^4 \, a_{2}(\eta)
- \frac{12\delta^5 }{5\pi} \, a_{1}(\eta)
\nonum \\
&\qquad ~~~+\frac{\delta^6 }{5} \, a_{0}(\eta)\, .
\label{axi16.1}
\end{flalign}
\end{subequations}
The general formula for $a_{2r}(\xi)$ has the form:
%
\be
a_{2r}(\xi) = a_{2r}^{(1)}(\xi) + a_{2r}^{(2)}(\xi) \, ,
\label{axi:a2s.1}
\ee
because $a_{2r}^{(3)}(\xi)=0$.

Importantly, the coefficient $a_{1}(\xi)$ cannot be excluded, because if $a_{1}(\xi_0)=0$ in some $\xi_0$-gauge
{(actually, we do have $a_{1}=0$ in the Landau gauge)},
it recovers a non-zero value in another gauge by the transformation for $a_{1}(\xi)$, Eq.~(\ref{sc:axi1}).
In a sense, the coefficient $a_{1}(\xi)$ (really, $a_{1}(\xi)/\pi$) behaves in a similar way to the even coefficients $a_{2r}(\xi)$.\\

{With the purpose of {checking} 
  the statement $a_{2m+1}(\xi)=0$ for $(m\geq 1)$ for $d=3$, we plan to perform a direct calculation of the
  coefficient $a_{3}$ in the Feynman and/or Landau gauge. Moreover, with the help of modern methods of calculations
  (see [\onlinecite{Kotikov:2018wxe}] for a recent review), we may calculate exactly the $\ep$-dependence of $a_{k}(\xi)$
  $(k=1,2,3)$ in $d=3-2\ep$ (or, at least, obtain the first few coefficients in the expansion with respect
  to $\ep$) and compare it
  with the $\ep$-dependence coming from LKF transformation, see Eqs.~(\ref{Pxid3}) and (\ref{Pxi.1d3}).}

\section{Summary and Conclusion}
\label{Sec:Conclusion}

In this work we have studied the LKF transformation for the massless fermion propagator of three-dimensional QED in the quenched approximation to all 
orders in the coupling $\alpha$.\,\footnote{{For recent studies of the LKF transformation in scalar QED$_3$, see \cite{Shubert2017}.}} 
Previous studies in
the literature were limited to the order $\alpha^2$. Our investigations were performed in dimensional regularization in $d=3-2\ep$ Euclidean space.

We have formulated two equivalent transformations: Eq.~(\ref{Pxid3}) together with (\ref{am.xid3}) and (\ref{hPhi})
on the one hand, and Eq.~(\ref{Pxi.1d3}) together with (\ref{axi:QED3}) and (\ref{tPhi}) on the other hand. Moreover, for the
coefficients $\hat{a}_m(\xi,p)$ in the transformation (\ref{Pxi.1d3}), we managed to obtain the closed expression (\ref{a_m-relation_final_explicit}) in terms
of the generalized Wright function ${}_1\Psi_1$ whose asymptotic expansion gives Eqs.~(\ref{axi:QED3}) and (\ref{tPhi}).

The transformation (\ref{Pxi.1d3}), which is similar to the ones used at lower orders in other papers
[\onlinecite{Bashir:2000rv,Bashir:2000iq,Bashir:2002sp,Bashir:2007zza}], allowed us to study the gauge-evolution
(from the $\eta$-gauge to the $\xi$-gauge) of each initial magnitude $a_m(\eta)$ and {to} reproduce
all results of the previous studies [\onlinecite{Bashir:2000rv,Bashir:2000iq}].

The other transformation, (\ref{Pxid3}), relates the magnitudes $a_m(\xi)$ in $\xi$-gauge to a combination
of initial magnitudes $a_l(\eta)$, where $0\leq l \leq m$. Studying this relation in dimensional regularization,
we observed that the contributions of odd magnitudes $a_{2t+1}(\eta)$ $(1\geq t \geq s-1)$ to even magnitudes
$a_{2s}(\xi)$ are accompanied by singularities which look like $\ep^{-1}$ in dimensional regularization, see Eq.~(\ref{axi:a2s3}).
In turn, the even magnitudes $a_{2s}(\eta)$ produce contributions to odd magnitudes $a_{2t+1}(\xi)$ $(t\geq s)$
$\sim \ep$ if $t\geq1$, see Eq.~(\ref{axi:a2s11}).

There are arguments in favor of ultraviolet and infrared perturbative finiteness of massless quenched QED$_3$ \cite{Jackiw1981,Karthik2017}.
Hence, assuming the existence of a finite limit as  $\ep \to 0$, we find that, exactly in $d=3$,
{\it all odd terms $a_{2t+1}(\xi)$ in perturbation theory, except $a_{1}$, should be exactly zero in any gauge}, \ie, even in the Landau gauge.

This statement is very strong and needs a further check. At the order $\alpha^2$, analytical expressions for the fermion self-energy diagrams
are well known. However, to the best of our knowledge, such results are absent at 
 three-loop order. We plan to study the $a_{3}$ term, \ie, three-loop diagrams, directly in the framework of perturbation theory in our future investigations.

Moreover, in our future studies, we also plan to consider the LKF transformation in unquenched QED$_3$ as well as
in the 
 {large-$N$ limit of QED$_3$} (see Refs. [\onlinecite{Gusynin:2016som,Kotikov:2016wrb,Kotikov:2016prf}]
and the recent review [\onlinecite{Kotikov:2020slw}]). In particular, this latter study will be performed in a non-local
gauge where we plan to apply results from previous studies [\onlinecite{Gorbar:2001qt,Teber:2012de,Teber:2014hna}] and
[\onlinecite{Ahmad:2016dsb,James:2019ctc}] in reduced QED$_{4,3}$ which is similar (see Ref. [\onlinecite{Kotikov:2016yrn}]) to QED$_3$
in the $1/N$-expansion.

\acknowledgments
We thank A.\ L.~Kataev for useful discussions.
The work of V.P.G.\ is supported by the National Academy of Sciences of Ukraine (project 0116U003191) and by its Program of Fundamental Research of
the Department of Physics and Astronomy (project No. 0117U000240).

\appendix


\section{Derivation of Eqs.~(\ref{am.xid3}) and (\ref{axi:QED3}) in a Euclidean space of dimension $d$ }
\label{sec:App:A}

In this Appendix, we shall present the LKF transformation in
a Euclidean space of dimension $d$. In the course of the evaluation, we shall also use the representation
$d=4-2\oep$, which is natural in four-dimensional space. Such a representation is useful because
the derived expressions are rather compact when expressed in terms of the $\oep$-dependence.
Accordingly, we do not make any decomposition in $\oep$ and, therefore, at any stage of the calculation in this Appendix,
all results have an exact $d$-dependence, when $\oep$ is replaced by $d$ as $\oep=(4-d)/2$.

Consider the general forms (\ref{SFp}) and (\ref{SFx}) of the fermion propagators $S_F(p,\xi)$ and $S_F(x,\xi)$,
in some gauge $\xi$. The latter are related  by the Fourier transforms (\ref{SFx2p}) and (\ref{SFp2x}). Moreover,
following Refs.~\cite{Gusynin1999PRD} and \cite{Kotikov:2019bqo,James:2019ctc} the propagators $S_F(x,\xi)$ and
$S_F(x,\eta)$ are related by the LKF transformation (\ref{def:LKFd}), where
\be
D(x) = e_d^2\,\Delta\,\mu^{2\oep}\,\int \frac{\D^d q}{(2\pi)^d } \, \frac{e^{-\I q x}}{q^4} \, .
\label{def:D(x):QEDd}
\ee


Using the Fourier transform (\ref{Fourier1}), we may evaluate $D(x)$ (\ref{def:D(x):QEDd}),
which is expressed as
\begin{flalign}
D(x) &= \bar{\al}_d  \,\Delta \,\mu^{2\oep}\,\Gamma(d/2-2)\,(\pi\,\mu^2\,x^2)^{2-d/2}
\nonum \\
&=
- \frac{\bar{\al}_d\,\Delta}{\oep}\,\Gamma(1-\oep)\,(\pi \mu^2 x^2)^{\oep}, \quad \bar{\al}_d = \frac{e_d^2}{16\pi^2}\, .
\label{DxQEDd}
\end{flalign}
Following standard practice, for dimensionally regularized calculations
 in a momentum space $d=4-2\oep$,
every loop produces additional factors $(4\pi)^{\oep}$ and $(\mu^2/p^2)^{\oep}$.
Then, assuming that the fermion propagator $S_F(p,\eta)$ with external momentum $p$ in some gauge $\eta$ takes
the form (\ref{SFp}), the function $P(p,\eta)$ can be written as {(see Refs.~\cite{Kotikov:2019bqo,James:2019ctc}).}
\be
P(p,\eta) = \sum_{m=0}^{\infty} a_m(\eta)\, \bar{\al}_d^m \,
{\left(\frac{\tilde{\mu}^2}{p^2}\right)}^{m\oep} \, ,
\label{Peta:QEDd}
\ee
where $a_m(\eta)$ are coefficients of the loop expansion of the propagator and
$\tilde{\mu}$ is the scale displayed in Eq.~(\ref{Aem})
which is intermediate between the MS scale $\mu$ and the $\overline{\rm{MS}}$ scale $\overline{\mu}$.

\begin{widetext}

In the following, we will derive exact formulas for the transformation $a_m(\eta) \to a_m(\xi)$ following
the LKF transformation (\ref{def:LKFd}) which is compact in $x$-space. In order to do so, it is convenient to
first derive an expression of $S_F(x,\eta)$ based on the ansatz (\ref{Peta:QEDd}) for $P(p,\eta)$.
Using the Fourier transform (\ref{Fourier1}), we have
\be
\int \frac{\D^d q}{(2\pi)^d } \, \frac{e^{-\I q x}}{(q^2)^{\beta}}  \, q_{\mu} = \left(\frac{i\, \partial}{\partial
  x_{\mu}}\right) \, \int \frac{\D^d q}{(2\pi)^d } \, \frac{e^{-\I q x}}{(q^2)^{\beta}}
=
\frac{1}{(4\pi)^{d/2}} \, \frac{2^{2\tilde \beta+1} \, a_1(\beta)\, x_{\mu}}{i \, (x^2)^{\tilde \beta+1}} \, ,
\label{Fourier3}
\ee
where $a_n(\beta)$ is defined in (\ref{tbeta}). Then, using  (\ref{SFp2x}), we obtain that:
\be
S_F(x,\eta) = \frac{2^{d-1}\,\hat{x}}{(4\pi\, x^2)^{d/2}} \,
\sum_{m=0}^{\infty} b_m(\eta)\, \bar{\al}_d^m \,
{\left(\pi\mu^2x^2\right)}^{m\oep} \, ,~~
b_m(\eta) =  a_m(\eta) \, \frac{\Gamma(d/2-m\oep)}{\Gamma(1+m\oep)} \, .
\label{bm.etad}
\ee

With the help of (\ref{bm.etad}) together with an expansion of the LKF exponent, we have

\be
S_F(x,\xi)=
S_F(x,\eta) e^{D(x)} = \frac{2^{d-1}\,\hat{x}}{(4\pi\, x^2)^{d/2}} \,\sum_{m=0}^{\infty} b_m(\eta)\, \bar{\al}_d^m \,
    {\left(\pi\mu^2x^2\right)}^{m\oep} \, \sum_{l=0}^{\infty}
 {\left(- \frac{\bar{\al}_d \,\Delta}{\oep}\right)}^l\,\frac{\Gamma^l(1-\oep)}{l!}\,(\pi \mu^2 x^2)^{l\oep}
\, .
\label{SFxid}
\ee
Factorizing all $x$-dependence yields:
\be
S_F(x,\xi) = \frac{2^{d-1}\,\hat{x}}{(4\pi\, x^2)^{d/2}} \,
\sum_{p=0}^{\infty} b_p(\xi)\, \bar{\al}_d^p \, {\left(\pi\mu^2x^2\right)}^{p\oep} \, ,~~
b_p(\xi) =  \sum_{m=0}^{p} \frac{b_m(\eta)}{(p-m)!}\, {\left(- \frac{\Delta}{\oep}\right)}^{p-m}\,\Gamma^{p-m}(1-\oep)
\, .
\label{bm.xid}
\ee

Hence, taking the correspondence between the results for propagators $P(p,\eta)$ and $S_F(x,\eta)$ in (\ref{Peta:QEDd}) and
(\ref{bm.etad}), respectively, together with the result (\ref{bm.xid}) for $S_F(x,\eta)$,
we have for $P(p,\xi)$:
\be
P(p,\xi) = \sum_{m=0}^{\infty} a_m(\xi)\, \bar{\al}_d^m \,
{\left(\frac{\tilde{\mu}^2}{p^2}\right)}^{m\oep} \, ,
\label{Pxid}
\ee
where
\be
a_m(\xi) = b_m(\xi) \, \frac{\Gamma(1+m\oep)}{\Gamma(d/2-m\oep)}
=  \sum_{l=0}^{m} \frac{a_l(\eta)}{(m-l)!} \,
\frac{\Gamma(d/2-l\oep)\Gamma(1+m\oep)}{\Gamma(1+l\oep)\Gamma(d/2-m\oep)}
 {\left(- \frac{\Delta}{\oep}\right)}^{m-l}\,\Gamma^{m-l}(1-\oep)\, .
\label{am.xid}
\ee
In this way, we have derived the expression of $a_m(\xi)$ using a simple expansion of the LKF exponent in $x$-space.
From this representation of the LKF transformation, we see that the magnitude $a_m(\xi)$ is determined by $a_l(\eta)$ with $0 \leq l \leq m$.

The corresponding result for the  $p$- and $\Delta$-dependencies of $\hat{a}_m(\xi,p)$
{(see definition of $\hat{a}_m(\xi,p)$ in Sec.~\ref{sec:LKF:x-space-analysis})}
can be obtained by interchanging the order in the sums in
the r.h.s.\ of (\ref{Pxid}). So, we have
\be
P(p,\xi) = \sum_{m=0}^{\infty} \hat{a}_m(\xi,p)\, \bar{\al}_d^m \,
{\left(\frac{\tilde{\mu}^2}{p^2}\right)}^{m\oep} \, ,
\label{Pxi.1d}
\ee
where
%
\be
  \hat{a}_m(\xi,p) = a_m(\eta) \,   \sum_{l=0}^{\infty}
  \frac{\Gamma(d/2-m\oep)\Gamma(1+(l+m)\oep}{\Gamma(1+m\oep)\Gamma(d/2-(l+m) \oep)}
  \,  {\left(- \frac{\bar{\al}_d\,\Delta}{\oep}\right)}^{l}\,\frac{\Gamma^{l}(1-\oep)}{l!}\,
       {\left(\frac{\tilde{\mu}^2}{p^2}\right)}^{l\oep} \, .
\label{ham.xid}
\ee

We would like to note that all of the above results may be expressed in $d=3-2\ep$ with the help of the substitutions $\bar{\ep} = 1/2 + \ep$
and $e_d^2 \,\mu = e^2$. The last replacement can also be expressed as  $\bar{\al}_d \, \mu = \al/(4\pi)$, with the
dimensionful $\al = e^2/(4\pi)$ defined in (\ref{alpha}).

  \section{Direct derivation of Eq.~(\ref{axi:QED3})}
\label{sec:App:B}

The LKF transformation in momentum space is given by Eq.~(\ref{LKF:Fourier}) where
\be
G(p,\Delta) = \int \D^d x\,e^{\I p x}e^{-b(x^2)^\nu}\, ,
\label{G(p):def}
\ee
with positive $b$ and $\nu$ from Eq.~(\ref{Mellin-int-D}).
In what follows we shall first compute (\ref{G(p):def}) and then consider the convolution integral (\ref{LKF:Fourier}).
\\

{\bf 1.}~Using the Mellin representation for the exponent $e^{-b(x^2)^\nu}$, Eq.~(\ref{Mellin-int-D}), we apply Eq.~(\ref{Fourier2})
to integrate over $x$. Performing the calculations, we obtain
\be
G(p,\Delta) = \left(\pi b^{-1/\nu}\right)^{d/2}\frac{1}{2\pi \I}\int\limits_{\gamma-\I \infty}^{\gamma + \I\infty}\D s\,\left(\frac{4b^{1/\nu}}{p^2}
\right)^{d/2-\nu s}\frac{\Gamma(s)\Gamma(d/2-\nu s)}{\Gamma(\nu s)}\, ,
\label{MB-for-G}
\ee
where the contour separates the poles of gamma functions in the numerator. Making the change of the variable, $s\to(d/2-s)/\nu$,
Eq.~(\ref{MB-for-G}) can be brought to the form:
\be
G(p,\Delta) = \left(\pi b^{-1/\nu}\right)^{d/2}{\cal E}_{\nu,d/2}\left(-\frac{p^2}{4b^{1/\nu}}\right)\, ,
\label{G-through-E}
\ee
which corresponds to Eq.~(\ref{G(p):exact}) in the main text. Here the two-parameter function ${\cal E}_{\nu,\alpha}(z)$ is defined by
the Mellin-Barnes integral in the complex plane:
\be
{\cal E}_{\nu,\alpha}(-z)=\frac{1}{2\pi \I}\int\limits_{\gamma-\I\infty}^{\gamma+\I\infty}
\D s z^{-s}\frac{\Gamma(s)\Gamma\left(\frac{\alpha-s}{\nu}\right)}{\nu\Gamma(\alpha-s)},\quad 0<{\Re}(\gamma)<{\Re}(\alpha)\, .
\label{MBintegral-GEF-contourL}
\ee
According to Eqs.~(1.19.1), (1.19.2) in \cite{Bateman1} the integral converges absolutely in the region $|\mbox{arg}\, z|<\frac{\pi}{2\nu}$ and
defines a function analytical in that sector for $\nu>0$. The function $\Gamma(s)$ has a sequence of poles at
$s=-m, m=0,1,2,\dots$ with residues $(-1)^m/m!$. Calculating the residues at the poles $s=-m$ yields the series:
\be
{\cal E}_{\nu,\alpha}(z)=\frac{1}{\nu}\sum\limits_{m=0}^\infty\frac{z^m}{m!}\frac{\Gamma\left(\frac{\alpha+m}{\nu}\right)}{\Gamma(\alpha+m)}
=\frac{1}{\nu}{}_1\Psi_1\left(\frac{\alpha}{\nu},\frac{1}{\nu};\alpha,1;z\right)\, .
\label{gen_exp_function}
\ee
The function ${\cal E}_{\nu,\alpha}(-z)$ is a special case of the generalized Wright function ${}_1\Psi_1$,
and is one of generalizations of the classical Mittag-Leffler function \cite{Srivastava2009} (sometimes also called the generalized exponential function
\cite{Barvinsky2019}). One may check that, for $d=3$,
\be
{\cal E}_{\frac{1}{2},\frac{3}{2}}(-z)=\frac{8}{\sqrt{\pi}(1+4z)^2}
\ee
and $G(p,\Delta)$ from (\ref{G-through-E}) reduces to Eq.~(\ref{G-to-delta}).

Displacing the integration path in Eq.~(\ref{MBintegral-GEF-contourL}) parallel to the imaginary axis to the right,
we cross the poles of $\Gamma\left(\frac{\alpha-s}{\nu}\right)$ at $s=\alpha+\nu m$, $m=1,2,\dots$ and
produce an expansion in descending powers of the variable z. When the path is displaced over a finite number
of poles, the remainder term is given by the integral (\ref{MBintegral-GEF-contourL}) taken over the displaced path.
Thus, we obtain the following the power-like asymptotic expansion (see the discussion in Sec.III of Ref.~\cite{Barvinsky2019}):
\be
{\cal E}_{\nu,\alpha}(-z)=z^{-\alpha}\sum\limits_{m=1}^\infty\frac{\Gamma(\alpha+\nu m)}{\Gamma(-\nu m)}\frac{(-z^{-\nu})^m}{m!}+{\cal O}(z^{-\infty})\, .
\ee
The symbol ${\cal O}(z^{-\infty})$ means terms which decrease faster than any power of $z$, \ie, in an exponential manner.
 This yields a series representation for $G(p,\Delta)$ at large momenta:
\begin{flalign}
G(p,\Delta) = \left(\frac{4\pi}{p^2}\right)^{d/2}\,\sum_{l=1}^\infty
\frac{\Gamma(d/2 + l\,(1/2+\ep))}{l!\Gamma(-l\,(1/2+\ep))}\,\left(-K\right)^l\,
\left(\frac{\tilde{\mu}^2}{p^2}\right)^{l\,(1/2+\ep)}\, .
\label{G(p):series}
\end{flalign}
\\

{\bf 2.}~ We now compute the convolution integral (\ref{LKF:Fourier}) which translates into the relation for the function $P$:
\be
P(p,\xi)=\int\frac{\D^dk}{(2\pi)^d}\frac{p\cdot k}{k^2}P(k,\eta)G(p-k,\Delta)\, .
\ee
With the help of the ansatz (\ref{Peta:QED3}) and Eq.~(\ref{G-through-E}) together with the Mellin-Barnes
representation (\ref{MBintegral-GEF-contourL}) for
the function ${\cal E}_{\nu,\alpha}(-z)$, this yields:
\bea
P(p,\xi)&=&\left(\pi b^{-1/\nu}\right)^{d/2}\sum\limits_{m=0}^\infty a_m(\eta)\left(\frac{\alpha}{2\sqrt{\pi}}\right)^m \tilde{\mu}^{2m\epsilon}
\frac{1}{2\pi \I}\int\limits_{\gamma-\I \infty}^{\gamma+ \I \infty} \D s \left(4b^{1/\nu}\right)^{s}\frac{\Gamma(s)\Gamma\left(\frac{d/2-s}{\nu}\right)}
{\nu\Gamma(d/2-s)}\nonumber\\
&\times&\int\frac{\D^dk}{(2\pi)^d}\frac{p\cdot k}{(k^2)^{1+m\nu}[(p-k)^2]^s}\, .
\eea
The resulting momentum integral is of the simple massless-propagator type with numerator and can be computed with the help of
(see, \eg, Ref.~[\onlinecite{Kotikov:2018wxe}]):
\be
\int \frac{\D^d k}{(2\pi)^d}\, \frac{k^{\mu_1 \cdots  \mu_n}}{k^{2\al} (p-k)^{2\beta}}
= \frac{(p^2)^{d/2 - \al - \beta}}{(4\pi)^{d/2}}\,p^{\mu_1 \cdots \mu_n}\,G^{(n,0)}(\al,\beta)\, ,
\label{def:one-loop-p-int+tp}
\ee
where $k^{\mu_1 \cdots  \mu_n}$ denotes the traceless symmetric tensor,  and
\be
G^{(n,0)}(\al,\beta) = \frac{a_n(\al) a_0(\beta)}{a_n(\al + \beta -d/2)}, \qquad a_n(\al) = \frac{\Gamma(n+d/2 - \al)}{\Gamma(\al)}\, .
\label{one-loop-Gn}
\ee
Performing the integral yields:
\be
\int \frac{\D^d k}{(2\pi)^d}\, \frac{p\cdot k}{k^{2(1+m\nu)} (p-k)^{2s}}=\frac{(p^2)^{d/2-m\nu-s}}{(4\pi)^{d/2}}G^{(1,0)}(1+m\nu,s)\, .
\ee
Hence we obtain
\bea
P(p,\xi)=\sum\limits_{m=0}^\infty a_m(\eta)\left(\frac{\alpha}{2\sqrt{\pi}p}\right)^m\left(\frac{\tilde{\mu}}{p^2}\right)^{m\epsilon}
\frac{1}{2\pi \I}\int\limits_{\gamma-\I\infty}^{\gamma+\I\infty}
\D s \left(\frac{p^2}{4b^{1/\nu}}\right)^{d/2-s}\frac{\Gamma(s)\Gamma\left(\frac{d/2-s}{\nu}\right)}
{\nu\Gamma(d/2-s)}G^{(1,0)}(1+m\nu,s)\, .
\eea
The last series has the form (\ref{Pxi.1d3}) and, for the coefficients $\hat{a}_m(\xi,p)$, we obtain the expression
\bea
\hat{a}_m(\xi,p)= a_m(\eta)\, \frac{\Gamma(d/2-m\nu)}{\Gamma(1+m\nu)} \,\frac{1}{2\pi \I}
\int\limits_{\gamma-\I\infty}^{\gamma+\I\infty} \D s \left(\frac{p^2}{4b^{1/\nu}}\right)^{d/2-s}\frac{\Gamma\left(\frac{d/2-s}{\nu}\right)\Gamma(1+m\nu+s-d/2)}
{\nu\Gamma(d-m\nu-s)}\, ,
\eea
which after the change of the variable, $s\to s+d/2-1- m\nu$, can be written in terms of the generalized Wright function
\bea
\hat{a}_m(\xi,p)=  a_m(\eta)\,
\frac{\Gamma(d/2-m\nu)}{\nu\Gamma(1+m\nu)}\left(\frac{p^2}{4b^{1/\nu}}\right)^{1+m\nu}
{}_1\Psi_1\left[-\frac{p^2}{4b^{1/\nu}}\Bigr|\begin{array}{c}(m+\frac{1}{\nu},\frac{1}{\nu})\\ (\frac{d}{2}+1,1)\end{array}\right]\, .
\label{a_m-through-Wright}
\eea
Here we used the Mellin-Barnes representation for this function
\bea
{}_1\Psi_1\left[z\Bigr|\begin{array}{c}(a,A)\\ (b,B)\end{array}\right] =
\frac{1}{2\pi \I}\int \limits_{\gamma - \I \infty}^{\gamma + \I \infty} \D s\,(-z)^{-s}\frac{\Gamma(s)\Gamma(a-A s)}{\Gamma(b-B s)}\, ,
\eea
where the contour separates the poles of gamma functions in the numerator. Eq.~(\ref{a_m-through-Wright}) is Eq.~(\ref{a_m-relation_final_explicit})
from the main text.

The Wright functions ${}_p\Psi_q(z)$ are studied rather well in the literature, see, for example,
\cite{Kilbas2002,Srivastava2009,Paris2010}.
Their series expansions can be derived from
Mellin-Barnes representation. For ${}_1\Psi_1$ deforming the contour to the left we obtain the small $z$ expansion,
\be
{}_1\Psi_1\left[z\Bigr|\begin{array}{c}(a,A)\\ (b,B)\end{array}\right]=\sum\limits_{k=0}^\infty\frac{\Gamma(a+A k)}{\Gamma(b+B k)}\frac{z^k}{k!}\, .
\label{Wright-small-z-exp}
\ee
On the other hand, deforming the contour to the right one going from $\infty-i\delta$ to $\infty+i\delta$ and enclosing the poles of $\Gamma(a-A s)$,
we can evaluate the residues at $s=\frac{l+a}{A},l=0,1,\dots$ and obtain the asymptotic expansion at large $z\gg1$,
\bea
{}_1\Psi_1\left[z\Bigr|\begin{array}{c}(a,A)\\ (b,B)\end{array}\right]=A^{-1}(-z)^{-a/A}\sum\limits_{l=0}^\infty\frac{(-1)^l}{l!}
\frac{\Gamma\left(\frac{a+l}{A}\right)}{\Gamma\left(b-B\frac{a+l}{A}\right)}(-z)^{-l/A}\, .
\eea
Correspondingly, the relation between the coefficients $\hat{a}_m(\xi,p)$ and $a_m(\eta)$ takes the following form at large momenta
\be
\hat{a}_m(\xi,p)=a_m(\eta)\frac{\Gamma(d/2-m\nu)}{\Gamma(1+m\nu)}\sum\limits_{l=0}^\infty\frac{(-K)^l}{l!}\frac{\Gamma(1+(m+l)\nu)}
{\Gamma(d/2-(m+l)\nu)}\left(\frac{\tilde{\mu}^2}{p^2}\right)^{l\nu}\, ,
\label{a_m-relation_fin}
\ee
which corresponds to Eq.~(\ref{axi:QED3}) in the main text.

For $d=3$ ($\epsilon=0$) the function ${}_1\Psi_1$ in Eq.~(\ref{a_m-relation_final_explicit}) takes the form
\bea
{}_1\Psi_1\left[-\left(\frac{p}{\alpha\Delta}\right)^2\Bigr|\begin{array}{c}(m+2,2)\\ (5/2,1)\end{array}\right]=\frac{4}{3\sqrt{\pi}}
\Gamma(m+2){}_2F_1\left(\frac{m+2}{2},\frac{m+3}{2};\frac{5}{2};-\left(\frac{2p}{\alpha\Delta}\right)^2\right)\, ,
\label{1Psi1-d=3}
\eea
where we used the series representation (\ref{Wright-small-z-exp}) and the duplication formula for the gamma function.
Eqs.~(\ref{1Psi1-d=3}) and (\ref{a_m-relation_final_explicit}) reproduce Eq.~(\ref{axi.ep=0-tot2}) in the main text.

Interestingly, Eq.~(\ref{1Psi1-d=3}) is finite and so the singularities appearing in (\ref{axi.ep=0-tot2})
are entirely due to the first gamma function in the numerator of (\ref{a_m-through-Wright}).
From the latter, we see that these singularities only affect odd coefficients with $m=2s+1$ and $s\geq 1$.
These singularities can be regularized with the help of a leading order in $\ep$ expansion. To this end,
it is enough to keep $\ep \not=0$ in the singular gamma function and set $\ep=0$ in all other terms. This yields:
\bea
\hat{a}_m(\xi,p)=  a_m(\eta)\,
\frac{8}{3\sqrt{\pi}}\,\frac{\Gamma((3-m)/2-(1+m)\ep)\,\Gamma(2+m)}{\Gamma(1+m/2)}\,\left(\frac{p}{\al \Delta}\right)^{2+m}\,
{}_2F_1\left(\frac{m+2}{2},\frac{m+3}{2};\frac{5}{2};-\left(\frac{2p}{\alpha\Delta}\right)^2\right)\, .
\label{a_m-through-Wright.exp1}
\eea
%

For even $m=2s$ the coefficients $\hat{a}_{2s}(\xi,p)$ are finite when $\ep=0$, they are obtained from Eq.~(\ref{a_m-through-Wright.exp1})
 and  take the form
\be
\hat{a}_{2s}(\xi,p)=  a_{2s}(\eta)\,
\frac{(-1)^s\,(1-4s^2)}{3}\left(\frac{2p}{\alpha\Delta}\right)^{2(1+s)}\,
{}_2F_1\left(s+1,s+\frac{3}{2};\frac{5}{2};-\left(\frac{2p}{\alpha\Delta}\right)^2\ \right)\, ,
\label{a_m-through-Wright.exp2.even1}
\ee
in accordance with Eq.~(\ref{axi.ep=0-tot2}).
The odd coefficients $\hat{a}_{2s+1}(\xi,p)$ are singular for $s\ge1$. Their leading behavior when $\epsilon\to0$ is easily obtained
from Eq.~(\ref{a_m-through-Wright.exp1}) and gives Eq.~(\ref{ha2s1:odd}) in the main text.

\section{Evaluation of the coefficients $a_k(\xi)$ at $\ep \ra 0$}
\label{sec:App:C}

As it was shown in subsection \ref{sec:LKF2a}, the accurate
analysis of the coefficients $a_k(\xi)$ requires considering the cases of even and odd values of $k$ separately.\\

{\bf 1}~ In the case where $k=2r$, we see from Eq.~(\ref{hPhi}) that:
\be
\hat{\Phi}(m,2r,\ep)= \frac{\Gamma(3/2-m/2-(m+1)\ep)\Gamma(1+r+2r\ep)}{\Gamma(1+m/2+m\ep)\Gamma(3/2-r-(2r+1)\ep)} \, .
\label{hPhi:even}
\ee
With the help of this expression, we proceed on considering even and odd values of $m$ separately.\\

{\bf 1.1}~ Let $m=2s$, then:
\be
\hat{\Phi}(2s,2r,\ep) =
\frac{\Gamma(3/2-s-(2s+1)\ep)\Gamma(1+r+2r\ep)}{\Gamma(1+s+2s\ep)\Gamma(3/2-r-(2r+1)\ep)}
\stackrel{\ep=0}{=} \frac{\Gamma(3/2-s)\Gamma(1+r)}{\Gamma(1+s))\Gamma(3/2-r)}\, .
\label{hPhi:even,even}
\ee
%
With the help of the
useful relation
\be
\frac{\Gamma(1-a-r)}{\Gamma(1-a)}=\frac{(-1)^r\Gamma(a)}{\Gamma(a+r)} \, ,
\label{Usrel}
\ee
Eq.~(\ref{hPhi:even,even}) can be brought to the form:
\be
  \hat{\Phi}(2s,2r,\ep)
  \stackrel{\ep=0}{=} (-1)^{r+s} \, \frac{\Gamma(r-1/2)\Gamma(1+r)}{\Gamma(1+s))\Gamma(s-1/2)}
  \, .
\label{hPhi:even,even.2}
\ee
The corresponding values of $\phi(2r-2s,0)$
%
\be
\phi(2r-2s,0) =
\frac{\pi^{r-s}}{[2(r-s)]!}
= \frac{\pi^{r-s} \Gamma(1/2)}{2^{2(r-s)}\Gamma(r-s+1/2)(r-s)!}\, ,
\label{phi:2r2s}
\ee
%
lead to
%
\be
\phi(2r-2s,0) \, (-2\Delta)^{2(r-s)} =
\frac{\Gamma(1/2)}{\Gamma(r-s+1/2)(r-s)!} \, (-\delta)^{2(r-s)}
\, ,
\label{delta}
\ee
%
{where $\delta$ was defined after Eq.~(\ref{axi:a2s3}).}
So, at the end, we have {Eq.~(\ref{axi:a2s1}) in the main text.}\\

{\bf 1.2}~ Let $m=2s+1$, then
%
\be
\hat{\Phi}(2s+1,2r,\ep) =
\frac{\Gamma(1-s-2(s+1)\ep)\Gamma(1+r+2r\ep)}{\Gamma(3/2+s+(2s+1)\ep)\Gamma(3/2-r-(2r+1)\ep)}
\label{hPhi:even,odd}
\ee
%
and we should consider the cases $s=0$ and  $s\geq 1$, separately.

In the case $s=0$, we have:
%
\be
\hat{\Phi}(1,2r,\ep)
\stackrel{\ep=0}{=} \frac{\Gamma(1+r)}{\Gamma(3/2))\Gamma(3/2-r)}
= 2 (-1)^{r+1}
 \frac{\Gamma(1+r)\Gamma(r-1/2)}{\pi^{3/2}}
 \, .
\label{hPhi:even,odd.2}
\ee
%
Since
%
\be
\phi(2r-1,0) \, (-2\Delta)^{2r-1}
= \frac{\pi^{r-1/2}  \, (-\Delta)^{2r-1}}{\Gamma(2r)}
= \frac{\Gamma(1/2) \pi^{r-1/2}  \, (-\Delta)^{2r-1}}{2^{2r-1}\Gamma(r)\Gamma(r+1/2)}
=\frac{\Gamma(1/2) \, (-\delta)^{2r-1}}{\Gamma(r)\Gamma(r+1/2)} \, ,
\label{phi2r1}
\ee
%
we have for $a_{2r}^{(2)}(\xi)$ {Eq.~(\ref{axi:a2sa1}) in the main text.}\\

For $s\geq 1$, the use of Eq.~(\ref{Usrel}) yields:
%
\be
\hat{\Phi}(2s+1,2r,\ep)= \frac{(-1)^{r+s+1}}{2(s+1)\ep} \,
\frac{\Gamma(1-2(s+1)\ep)\Gamma(1+2(s+1)\ep)}{\Gamma(s+2(s+1)\ep)\Gamma(3/2+s+(2s+1)\ep)} \,
\frac{\Gamma(r-1/2+(2r+1)\ep)\Gamma(1+r+2r\ep)}{\Gamma(1/2-(2r+1)\ep)\Gamma(1/2+(2r+1)\ep)}
  \, .
\label{hPhi:even,odd.3}
\ee
%
Taking the leading contribution at $\ep \to 0$,
we obtain:
\be
\hat{\Phi}(2s+1,2r,\ep) =
\frac{(-1)^{r+s+1}}{2(s+1)\pi \ep} \,
\frac{\Gamma(1+r)\Gamma(r-1/2)}{\Gamma(s)\Gamma(3/2+s)}  + O(\ep^0)  \, .
\label{hPhi:even,odd.4}
\ee
%
Since
\be
\phi(2r-2s-1,0) \, (-2\Delta)^{2r-2s-1} = \frac{\pi^{r-s-1/2}  \, (-\Delta)^{2r-2s-1}}{\Gamma(2r-2s)}
= \frac{\Gamma(1/2) \, (-\delta)^{2r-2s-1}}{\Gamma(r-s)\Gamma(r-s+1/2)} \, ,
\label{phi2rsm1}
\ee
%
we have for $a_{2r}^{(3)}(\xi)$ {Eq.~(\ref{axi:a2s3}) in the main text.}\\

{\bf 2.}~ In the case where $k=2r+1$, we see from Eq.~(\ref{hPhi})
that:
\be
\hat{\Phi}(m,2r+1,\ep) =
\frac{\Gamma(3/2-m/2-(m+1)\ep)\Gamma(3/2+r+(2r+1)\ep)}{\Gamma(1+m/2+m\ep)
    \Gamma(1-r-2(r+1)\ep)} \, .
\label{hPhi:odd}
\ee
%
It is convenient to consider separately the cases $r=0$ and $r\geq 1$.\\

{\bf 2.1}~ For $r=0$, we have:
\be
a_1^{(1)}(\xi) = \sum_{m=0}^{1} \, a_m(\eta) \,
\hat{\Phi}(m,1,\ep) \, \phi(1-m,\ep) \, (-2\Delta)^{1-m}
= a_0(\eta) \,
\hat{\Phi}(0,1,\ep) \, \frac{\Gamma(1/2-\ep)}{(1+2\ep)} \, (-2\Delta) + a_1(\eta) \, \hat{\Phi}(1,1,\ep) \, .
\label{axi:1}
\ee
%
Using (\ref{hPhi:odd}), we obtain
%
\be
\label{Phi11-Phi01}
\hat{\Phi}(1,1,\ep)=1\, ,~~
\hat{\Phi}(0,1,\ep)= \frac{\Gamma(3/2-\ep)\Gamma(3/2+\ep)}{\Gamma(1-2\ep)} = \frac{\pi}{4}
+ O(\ep^2)
\ee
%
and Eq.~(\ref{axi:1}) becomes {Eq.~(\ref{axi:1f}) in the main text.}\\

{\bf 2.2}~ For $r\geq 1$, using (\ref{hPhi:odd}),  we obtain
\be
  \hat{\Phi}(m,2r+1,\ep) =
  \frac{\Gamma(3/2-m/2-(m+1)\ep)}{\Gamma(1+m/2+m\ep)} \, (-1)^r [2(r+1)\ep] \,
  \frac{\Gamma(r+2(r+1)\ep)\Gamma(3/2+r+(2r+1)\ep)}{\Gamma(1-2(r+1)\ep)\Gamma(1+2(r+1)\ep)}
  \, .
\label{hPhi:odd.1}
\ee

In the case $m=2s$, we have
\begin{flalign}
  \hat{\Phi}(2s,2r+1,\ep) =
  \frac{\Gamma(3/2-s)}{\Gamma(1+s)} \, (-1)^r [2(r+1)\ep] \,
\Gamma(r)\Gamma(3/2+r) + O(\ep^2) \, .
\label{hPhi:oddeven}
\end{flalign}
The use of (\ref{Usrel}) then yields:
\begin{flalign}
  \hat{\Phi}(2s,2r+1,\ep)=
  (-1)^{s+r+1} [2\pi(r+1)\ep] \,
\frac{\Gamma(r)\Gamma(3/2+r)}{\Gamma(1+s)\Gamma(s-1/2)} + O(\ep^2)
  \, .
\label{hPhi:odd,even1}
\end{flalign}
Moreover, since
\begin{flalign}
  \phi(2r-2s+1,0) \, (-2\Delta)^{2r-2s+1} =
  \frac{\Gamma(1/2) \, (-\delta)^{2r-2s+1}}{\Gamma(r-s+1)\Gamma(r-s+3/2)} \, ,
\label{phi2rsp1}
\end{flalign}
we have (for $r \geq 1$) {Eq.~(\ref{axi:a2s11}) in the main text.}\\

In the case $m=2s+1$, we have
\begin{flalign}
  \hat{\Phi}(2s+1,2r+1,\ep) =
  \frac{\Gamma(1-s-2(s+1)\ep)}{\Gamma(3/2+s+(2s+1)\ep)} \, (-1)^r [2(r+1)\ep] \,
  \frac{\Gamma(r+2(r+1)\ep)\Gamma(3/2+r+(2r+1)\ep)}{\Gamma(1-2(r+1)\ep)\Gamma(1+2(r+1)\ep)}
  \, .
\label{hPhi:odd.2}
\end{flalign}
It is convenient to consider the cases $s=0$ and $s\geq 1$ separately.

For $s=0$, Eq.~(\ref{hPhi:odd.2}) simplifies as:
\begin{flalign}
  \hat{\Phi}(1,2r+1,\ep) =
  \frac{\Gamma(1-2\ep)}{\Gamma(3/2+\ep)} \, (-1)^r [4(r+1)\ep] \,
\Gamma(r)\Gamma(3/2+r) + O(\ep^2)
= (-1)^r [4(r+1)\ep] \, \frac{\Gamma(r)\Gamma(3/2+r)}{\sqrt{\pi}} + O(\ep^2)
  \, .
\label{hPhi:odd1}
\end{flalign}

For $s \geq 1$, the use of (\ref{Usrel}) yields:
\begin{flalign}
  \hat{\Phi}(2s+1,2r+1,\ep) =
  (-1)^{r+s} \, \frac{(r+1)}{(s+1)} \,
\frac{\Gamma(r)\Gamma(3/2+r)}{\Gamma(s)\Gamma(3/2+s)} + O(\ep^1)
  \, .
\label{hPhi:odds}
\end{flalign}

With the help of (\ref{phi2rsp1}), we obtain the final expression for  $a_{2r+1}^{(2)}(\xi)$
{and $a_{2r+1}^{(3)}(\xi)$ Eqs.~(\ref{axi:a2s12}) and (\ref{axi:a2s13})} in the main text.

\end{widetext}

\end{document}